\documentclass[namedreferences]{solarphysics}
\usepackage[optionalrh]{spr-sola-addons} 
\usepackage{graphicx}        
\usepackage{color}           
\usepackage{url}             

\begin{document}

\begin{article}

\begin{opening}

\title{The Measurement of Solar Diameter and Limb Darkening Function with the Eclipse Observations}

\author{A.~\surname{Raponi}$^{1}$\sep
        C.~\surname{Sigismondi}$^{2}$\sep
        K.~\surname{Guhl}$^{3}$\sep
        R.~\surname{Nugent}$^{4}$\sep
        A.~\surname{Tegtmeier}$^{5}$\sep
        }
\runningauthor{Raponi et al.}
\runningtitle{LDF}

   \institute{$^{1}$ Sapienza University of Rome, P.le Aldo Moro 5 00185, Roma (Italy) 
                     email: \url{andr.raponi@gmail.com} \\ 
              $^{2}$ Sapienza University of Rome; ICRA, International Center for Relativistic Astrophysics, P.le Aldo Moro 5 00185, Roma (Italy);
                     University of Nice-Sophia Antipolis (France); Istituto Ricerche Solari di Locarno (Switzerland); GPA Observatorio Nacional, 
                     Rio de Janeiro (Brasil).
                     email: \url{sigismondi@icra.it} \\
              $^{3}$ IOTA, International Occultation Timing Association, European Section and Archenhold Sternwarte, Alt-Treptow 1 D 12435 Berlin (Germany)
                     email: \url{kguhl@astw.de} \\
              $^{4}$ IOTA, International Occultation Timing Association, US Section
                     email: \url{rnugent@wt.net} \\ 
              $^{5}$ IOTA, International Occultation Timing Association, European Section
                     email: \url{andreas.tegtmeier@freenet.de} \\ 
             }

\begin{abstract}
The total solar irradiance varies over a solar cycle of 11 years and maybe over cycles of longer periods. Is the solar diameter variable over time too? A discussion of the solar diameter and its variations must be linked to the limb darkening function (LDF).
We introduce a new method to perform high resolution astrometry of the solar diameter from the ground, through the observations of eclipses, using the luminosity evolution of Baily's Bead and the profile of the lunar edge available from satellite data. This approach unifies the definition of solar limb with the inflection point of LDF for eclipses and drift-scan or heliometric methods.
The method proposed is applied for the videos of the eclipse on 15 January 2010 recorded in Uganda and in India. The result suggests reconsidering the evaluations of the historical eclipses observed with a naked eye.

\end{abstract}


\end{opening}

\maketitle

\section{The Variability of the Solar Parameters}

Despite the observation of the changing nature of the solar surface dating back to the 17th century, the idea of the immutability of the solar luminosity was abandoned only recently. The ``solar constant" was in fact the name that \inlinecite{Pouillet} gave to the total electromagnetic energy received per unit area per unit of time, at the mean Sun-Earth distance (1AU). Today we refer to this parameter more correctly as total solar irradiance (TSI). The difficulties in measuring the TSI through the ever-changing Earth atmosphere precluded a resolution of the question about its immutability prior to the space age. Today there is irrefutable evidence for variations of the TSI. Measurements made during more than two solar cycles show variability on different time-scales, ranging from minutes up to decades, or more. Despite the fact that collected data came from different instruments aboard different spacecraft, it has been possible to construct a homogeneous composite TSI time series, filling the measurement gaps and adjusted to an initial reference scale. The most prominent discovery of these space-based TSI measurements is a 0.1\% variability over the solar cycle (see Figure 1), values being higher during phases of maximum activity \cite{Pap}. 

\begin{figure}
\centerline{\includegraphics[width=0.6\textwidth,clip=]{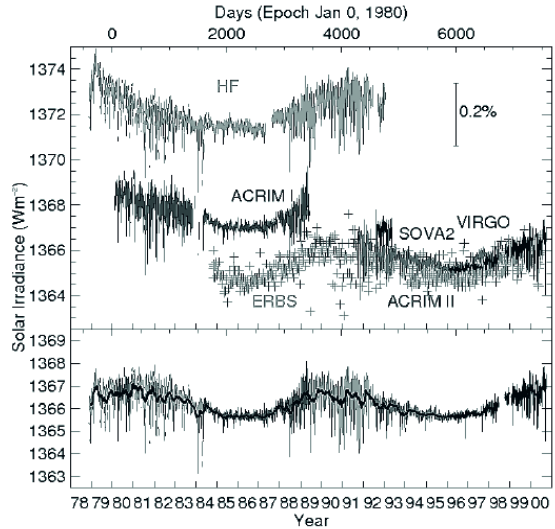}}
\caption{Various total irradiance time series are presented on the upper panel. The composite total solar irradiance is shown on the lower panel \protect\cite{Pap}.}
\label{Fig. 1}
\end{figure}

The link between solar activity and TSI has certainly paved the way for a deeper understanding of solar physics, and for a debate on the role of the Sun in Earth's climate. For example, the overlap between the Little Ice Age of the 17th century and the Maunder Minimum is a remarkable coincidence. 

Presently, the most successful models assume that surface magnetism is responsible for TSI changes on time-scales of days to years \cite{Solanki}. The modeled TSI versus TSI measurements made by the VIRGO experiment aboard the {\it Solar and Heliospheric Observatory} (SOHO), between January 1996 and September 2001, show a correlation coefficient of 0.96 \cite{Krivova}. However, significant variations in TSI remain unexplained after removing the effect of sunspots, faculae, and the magnetic network \cite{Pap,Kuhn}.
Other possible causes of the variation of the TSI may involve the global parameters of the Sun: temperature and radius\footnote[1]{Dealing with the radius is equivalent to dealing with the diameter, but in this work we refer to the former or to the latter depending on the context.}. 

Recently, monitoring the spectrum of the quiet-Sun atmosphere at the center of the solar disk during thirty years at Kitt Peak, \inlinecite{Livingston} have shown a "nearly immutable basal photosphere temperature" during the solar cycle within the observational accuracy, {\it i.e.} $\delta T=0 \pm 0.3$ K.

The solar radius is the global property with the most uncertain determination for the changes over a solar cycle. The most accurate measurements (with accuracy better than 0.1 arcsec) are still far from an agreement \cite{Djafer}. 
The lack of agreement between near simultaneous ground-based measurements at different locations suggests that atmospheric contamination is severe. Moreover the lack of coherence of the set of the values obtained from different observers can also be explained by the lack of a common strategy: different instrumental characteristics, different choices of wavelength, and different processing methods. The measurements from space are very few and they also have some sources of doubt. The Michelson Doppler Imager (MDI) on board the SOHO satellite indicates a change of no more than $23 \pm 9$ mas (milli-arcseconds) in phase with solar activity \cite{Bush}, while the Solar Disk Sextant (SDS; \citeauthor{Sofia}, \citeyear{Sofia}), during five flights between 1992 and 1996, measured a variation of 0.2 arcsec in anti-phase with respect to the solar activity \cite{Egidi}. Although milli-arcsecond sensitivity was achieved for the SDS, its results seem to be too large according to the studies on helioseismology, in particular to the $f$-mode frequencies \cite{Antia,Dziembowski}.

The variability of the solar diameter over periods longer than a solar cycle is even more enigmatic. \inlinecite{Ribes} reported the measurements performed at the Observatoire de Paris from 1660-1719 taken by French astronomers, including Jean Felix Picard, and the Italian director Giovanni Domenico Cassini. They measured a solar diameter $7\pm1$ arcsec larger than the standard value of 1919.26 arcsec \cite{Auwers} that is the modern reference for the absolute value of the solar diameter. As we will see in Section 5, this higher value is in a sense confirmed by the observation of historical eclipses. Although this strengthens the hypothesis of a larger diameter in the past, discussion on these absolute measurements must be made with care, as we are going to show.

\section{Observing the Limb Darkening Function}

Solar images in the visible wavelength range show that the disk center is brighter than the limb region. This phenomenon is known as the limb darkening, and is described by the limb darkening function (LDF). The position of the inflection point of the LDF is the conventionally accepted definition of the solar limb. In this article we adopt this definition.

\subsection{Seeing Effect}
In spite of the improvements in the measurements, there is always uncertainty resulting from the effect of Earth's atmosphere (seeing) and this effect is considered to be the main source of the discrepancies among the diameter determinations. The effect, as shown in Figure 2, is the smoothing of the steepness of the profile on the limb, moving inwards the inflection point. 

\begin{figure}
\centerline{\includegraphics[width=0.8\textwidth,clip=]{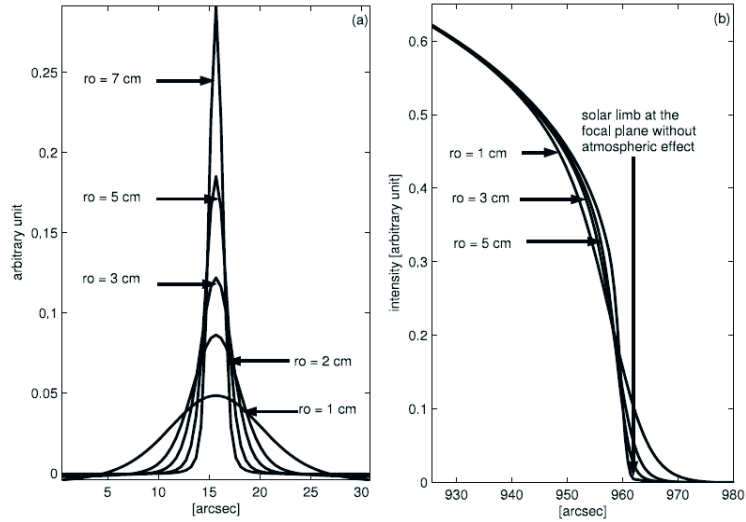}}
\caption{(a) Point spread functions for several values of Fried's parameter $r_{0}$ according to the Kolmogorov turbulence model, through a telescope having $D$ = 10 cm. (b) Effect of atmospheric turbulence on the solar limb \protect\cite{Djafer}.}
\label{Fig. 2}
\end{figure}

\subsection{Instrumental Effect}
Instrumental effect depends on the characteristics of each telescope. 

The full-width at half-maximum (FWHM) of the point spread function (PSF) is proportional to $\lambda/D$, where $\lambda$ is the wavelength of the observation and $D$ is the pupil diameter. 
This is similar to the FWHM of the atmospheric PSF represented in Figure 2, that is proportional to $\lambda/r_{0}$. Then for a given value of the wavelength, if $D$ decreases, the FWHM of the PSF increases, the inflection point moves inwards, and consequently the calculated diameter decreases (Figure 2b). Thus, two instruments with different $D$ will measure different solar diameters if this instrumental effect is not taken into account. 

\subsection{Dependence on Wavelength}
The position of the solar limb depends on wavelength not only through the PSF (instrumental or atmospheric) but also on the layer at which the specific wavelength observes.

The dependence of the solar limb position on the wavelength has a sign opposite to that seen for the PSF; longer wavelengths have the inflection point shift outwards. 

The effect of the presence of Fraunhofer lines is illustrated by \inlinecite{Thuillier}. The authors reconstructed the limb profile for different wavelengths, including the continuum and the center of spectral lines, with the Code for Solar Irradiance (COSI), developed by \inlinecite{Haberreiter} and improved by \inlinecite{Shapiro}. The results on the prediction of inflection point position is shown in Figure 3.

\begin{figure}
\centerline{\includegraphics[width=0.7\textwidth,clip=]{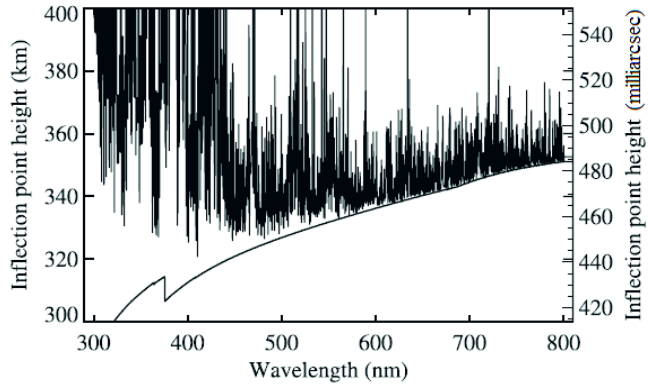}}
\caption{COSI model predictions of the position of the inflection point as a function of wavelength for the continuum (lower solid line) and for the case when Fraunhofer lines are taken into account (upper ``zig-zag" lines corresponding to the outer inflection point when observed with very narrow band-pass filters) \protect\cite{Thuillier}.}
\label{Fig. 3}
\end{figure}

\subsection{Asphericity}
The asphericity is the observed variability of the radius $R$ with the solar latitude. The asphericity of the Sun, and in particular the oblateness: $f=(R_{\rm eq}-R_{\rm pol})/R_{\rm eq}$ (`eq' and `pol' are for the equator and the pole), has great implications on the motion of the bodies around the Sun through the quadrupole moment \cite{Oliva}. It can also provide important information on solar physics as with the other global parameters.

Measuring the radius of the Sun for several solar latitudes may provide a measure on asphericity. Conversely, for monitoring the variation of the solar radius, one must take into account that different solar latitudes may give different radii because of the asphericity, but anyway not more than 20 mas according to \inlinecite{Rozelot}.

\subsection{Solar Surface Magnetic Structure}
The position of the inflection point also depends on the existence of active regions. This is discussed in detail by \inlinecite{Thuillier}. Because of the lack of observations on this topic, the authors make use of some models of the solar atmosphere. Despite the differences between the results of the models, a common trend is clear: The presence of faculae displaces the inflection point outward with respect to its location predicted by the quiet Sun models, while sunspots displace it inward.

The observation of the LDF in conjunction with these phenomena can help to discern between the models, providing useful information on the solar atmosphere. But if the goal of the measurement is the monitoring of the solar diameter, one has to avoid measuring the limb portions with active regions.

\section{The Method of Eclipses}
\subsection{Comparison between Observation and Ephemeris}
With the eclipse observation we are able to bypass some problems that affect the measurement of solar diameter. The atmospheric and instrumental effects that distort the shape of the limb (see Sections 2.1 and 2.2) are overcome by the fact that the scattering of the Sun's light is greatly reduced by the occultation of the Moon. Therefore there are much less photons from the photosphere to be poured, by the PSF effect, in the outer region. 

A decisive breakthrough in the determination of the solar diameter by the eclipse observation was made thanks to David Dunham who proposed to observe Baily's beads in connection with lunar profile data. 

Baily's beads, from \inlinecite{Baily} who first explained the phenomenon, are beads of light that appear or disappear from the bottom of a lunar valley when the solar limb is almost tangent to the lunar edge. 

It is not their positions to be directly measured, but the timing of appearance or disappearance. In fact, the times when the photosphere disappears or emerges behind the valleys of the lunar edge are determined solely by the positions of the solar and lunar limbs and the lunar profile involved at the instant, bypassing in this way the atmospheric seeing.

The International Occultation Timing Association (IOTA) is currently engaged to observe the eclipses with the aim of measuring the solar diameter. This is facilitated by the development of the software Occult 4 by David Herald\footnote[2]{\url{http://www.lunar-occultations.com/iota/occult4.htm}}.

The technique consists of looking at the time of appearance of the beads and comparing it with the calculated positions by the ephemeris using the software Occult 4 \cite{Sigi2009}. The simulated Sun by Occult 4 has the standard radius: 959.63 arcsec at 1AU \cite{Auwers}. The difference between the simulations and the observations is a measure of the radius correction with respect to the standard radius ($\Delta R$), within the accuracy on the ephemeris (few mas).

In the polar regions of the Moon, due to the geometry of the eclipse the beads last for longer time \cite{Sigi2009}. Observing at the polar regions we are also able to avoid the undesirable effect of solar active regions (see Section 2.5). In fact solar active regions do not appear in solar latitude higher than $\approx 40^\circ$, and the maximum offset between the lunar and the solar poles is $\approx 9^\circ$ in the axis angle (AA, by Occult 4 calculation). 

In grazing eclipses the number $N$ of the beads can be high, providing $N$ determinations of the photosphere's circle.

\begin{figure}
\centerline{\includegraphics[width=0.6\textwidth,clip=]{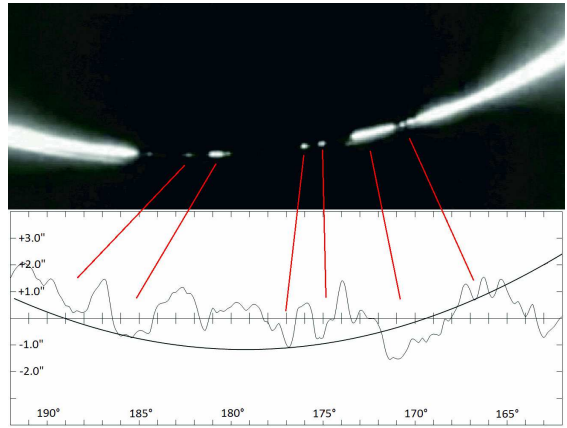}}
\caption{The upper panel shows an instant of the annular eclipse in Uganda (2$^{\circ}$ 41' 19.8" N, 32$^{\circ}$ 19' 2.9" E, 1092 m above sea level, 15 January 2010 05:25:48 UTC) filmed by Richard Nugent. The image is horizontally reversed with respect to the original video to fit it with the lower panel. The lower panel shows the lunar profile by {\it Kaguya} (low resolution) plotted by the Occult 4 software. In the abscissa the axis angle (AA) is the angle around the limb of the Moon, measured eastward from the Moon's north pole. The distance in arcsec from the mean lunar edge is in the ordinate. The curved line is the standard solar limb, which is the limb of the Sun with the standard radius and position by ephemeris calculation.}
\label{Fig. 4}
\end{figure}

However, this approach conceives the bead as an on-off signal. In this way one assumes the LDF as a Heaviside profile, but it is actually not. Different compositions of optical instruments (telescope + filter + detector) could have different sensitivities and different signal-to-noise ratios, recording the first signal of the bead in correspondence of different points along the luminosity profile. This leads to different values of $\Delta R$ (see Figure 5).
An improvement of this approach has to take into account the whole shape of the LDF and thus the actual position of the inflection point.

\begin{figure}
\centerline{\includegraphics[width=0.6\textwidth,clip=]{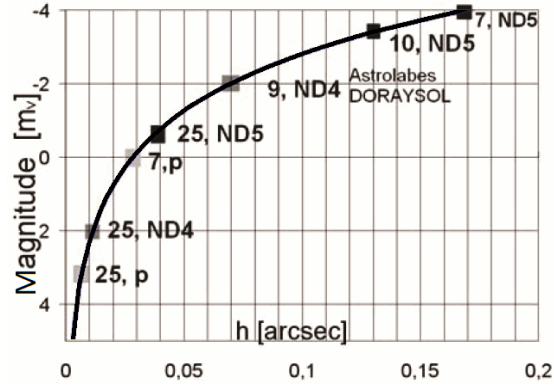}}
\caption{The evolution of the magnitude of the bead. The height of the solar limb above the valley is on the abscissa. Various squares represent different types of telescopes: [25, p] means 25 cm aperture with projection of the image, [25, ND4] the 25 cm telescope with a neutral density filter of transmittance 1/10000, ND5 stands for 1/100000, and so on.}
\label{Fig. 5}
\end{figure}

\subsection{Numerical Calculation}

The shape of the light curve of the bead is determined by the shape of the LDF (not affected by seeing) and the shape of the lunar valley that generates the bead. 
Calling $w(x)$ the width of the lunar valley ({\it i.e.} the length of the solar limb visible from the valley as a function of the height $x$ from the bottom of the valley), and $B(x)$ the surface brightness profile ({\it i.e.} the LDF), one could see the light curve $L(y)$ as a convolution of $B(x)$ and $w(x)$, being $\mid y\mid$ the distance between the bottom of the lunar valley and the standard solar limb (see Figure 4), setting to 0 the position of the standard limb (the point 0 is not the inflection point because its position is our goal):

\noindent $L(y)=\intop B(x)\, w(y-x)\, dx$

\noindent A discretized convolution can be written as:

\noindent $L(m)=\sum B(n)\, w(m-n) h$

\noindent where $n$ and $m$ are the indices of the discrete layers corresponding to the $(x, y)$ coordinates and $h$ is the layers thickness.

Thus the profile of the LDF is discretized in order to obtain the solar layers $B(n)$ of equal thickness and concentric to the center of the Sun. In the limited horizontal scale of a lunar valley these layers are regarded roughly parallel and straight. 

The lunar valley is also divided into layers of equal angular thickness. During a bead event every lunar layer is filled by one solar layer for every given interval of time. Step by step during an emerging bead event, a deeper layer of the solar atmosphere enters into the profile drawn by the lunar valleys (see Figure 6), and each layer casts light through the same geometrical area of the previous one. 

\begin{figure}
\centerline{\includegraphics[width=0.8\textwidth,clip=]{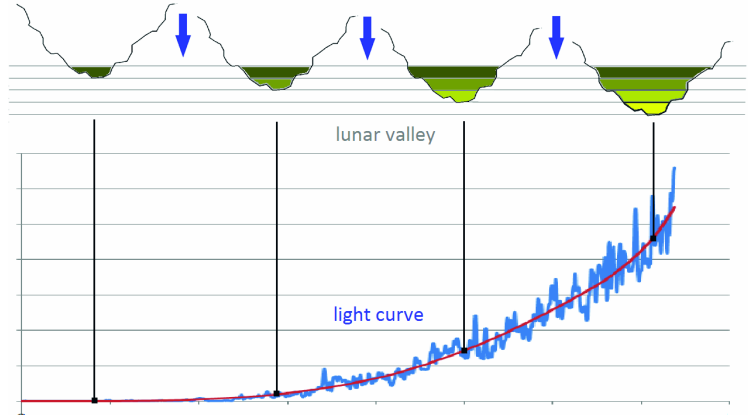}}
\caption{Every step in the geometry of the solar-lunar layers (upper panel) corresponds to a given instant in the light curve (lower panel). The light curve gives the contributions from all the layers.}
\label{Fig. 6}
\end{figure}

Let $A_{1},\, A_{2}..\: A_{N}$ be the area of lunar layers from the bottom of the valley going outward, $B_{1},\, B_{2}..\: B_{N}$ be the surface brightness of the solar layers (our goal) from the outer (dimmer) going inward, and $L_{1},\, L_{2}..\: L_{N}$ be the value of the observed light curve from the first signal to the saturation of the detector or to the replenishment of the lunar valley. One then obtains: 

\noindent $L_{m} = B_{1}A_{m} + B_{2}A_{m-1} + .. + B_{m}A_{1} = \sum_{n=1}^m B_{n}\, A_{m-n+1}$

To derive the LDF profile, one can proceed:

\noindent $B_{1}=L_{1}/A_{1}$

\noindent $B_{2}=\left[L_{2}-(B_{1}\cdot A_{2})\right]/A_{1}$

\noindent $B_{3}=\left[L_{3}-\left(B_{1}\cdot A_{3}+B_{2}\cdot A_{2}\right)\right]/A_{1}$

\noindent $B_{4}=\left[L_{4}-\left(B_{1}\cdot A_{4}+B_{2}\cdot A_{3}+B_{3}\cdot A_{2}\right)\right]/A_{1}$ 

\noindent and so on. 

\noindent The situation described above is relate to an emerging bead. The same process can be applied to a disappearing bead, simply plotting the light curve back in time.
The LDF obtained in this way is a discrete LDF. The smaller is the thickness of each layer, the higher is the resolution of the LDF.
The angular thickness of the layers, {\it i.e.} the sampling, depends on our knowledge about the error of the lunar profile. 

What we need for the deconvolution procedure is thus:
\begin{itemize}
\item{the light curve of the bead event,}
\item{the area of each lunar layer involved, and}
\item{the correspondence between $L(y)$ and $L(t)$, {\it i.e.} the motion of the solar limb in the lunar valley.}
\end{itemize}

\section{An Application of the Method of Eclipses}
We studied the videos of the annular eclipse on 15 January 2010 obtained by Richard Nugent in Uganda and Andreas Tegtmeier in India. 

The equipment of Nugent was: CCD camera (Watec 902H Ultimate)\footnote[3]{\url{http://www.aegis-elec.com/products/watec-902H_spec_eng.pdf}}; Matsukov telescope (90 mm aperture, 1300 mm focal length); panchromatic ND5 filter (Thousand Oaks). 

The equipment of Tegtmeier was: CCD camera (Watec 120N)\footnote[4]{\url{http://www.aegis-elec.com/products/watec-100N_spec_eng.pdf}}; Matsukov telescope (100 mm aperture, 1000 mm focal length); IOTA/ES green glass-based neutral ND4 filter.

\begin{figure}
\centerline{\includegraphics[width=0.8\textwidth,clip=]{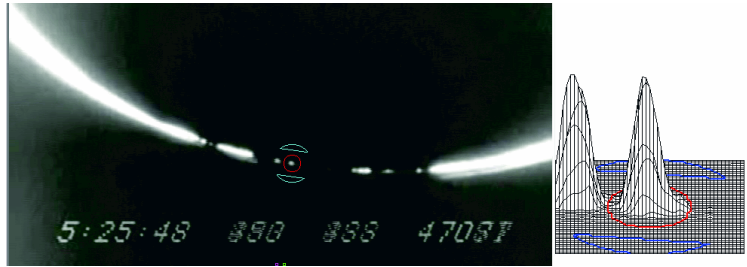}}
\caption{Limovie software. An instant of the light curve analysis is shown. The red circle selects the area where the light intensity is recorded as a function of time. The number of pixels involved in this area are $\pi r^{2}=380$, where $r$ is a settable radius of the red circle. The two blue sections of the circle select the area where the background noise is recorded. On the right panel a 3D graph shows the intensity of the bead. This is useful in the choice of the radius of the red circle and in the evaluation of the saturation of the CCD.}
\label{Fig. 7}
\end{figure}

\subsection{Light Curve Analysis}
Two beads located at AA $=171^\circ$ and $177^\circ$ are analyzed for the two videos. They were disappearing for Nugent's video and appearing for Tegtmeier's video. For the analysis of the light curve a software specially developed for this purpose is used: the Limovie free software\footnote[5]{\url{http://www005.upp.so-net.ne.jp/k_miyash/occ02/limovie_en.html}} (see Figure 7). 

The standard deviation $\sigma$ of the background noise is calculated. Then a constant value of $5\sigma$ is subtracted from the light curve. The light curves of the disappearing beads are plotted back in time. The first positive value is considered the first signal. The negative values are set equal to 0. Then some polynomial fits are performed in order to reduce the electronic noise (see Figure 8). 

\begin{figure}
\centerline{\includegraphics[width=1.0\textwidth,clip=]{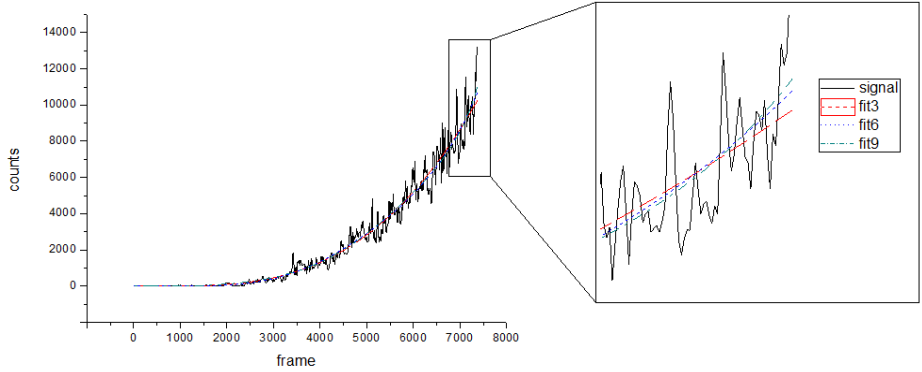}}
\caption{Light curve of the disappearing bead at AA = 177$^\circ$. Increasing frame numbers correspond to decreasing time. The velocity of acquisition is 30 frames per second. The counts on the y-axis are the sum of the intensities of 380 pixels involved in the bead. Each pixel has 256 levels of intensity (8 bits). A detail of the light curve is zoomed on the right-hand-side panel. The polynomial fits (fit3--fit9, for the third to ninth-order polynomials) are shown.}
\label{Fig8}
\end{figure}

\subsection{The Thickness of the Layer}
The lunar valley analysis is performed with the new lunar profile\footnote[6]{\url{http://wms.selene.jaxa.jp/selene_viewer/index_e.html}} obtained by the laser altimeter (LALT) onboard the Japanese lunar explorer {\it Kaguya} \cite{Araki}. The error in radial topography is estimated to be ± 4.1 m corresponding to 2.1-2.3 mas from the distance of the Earth.

The lunar valley has to be divided into layers as explained in Section 3 (see Figure 9). The layer at the bottom of the valley ($A_1$) is special, because (i) it defines the thickness of all the other layers as well, (ii) its area, being most heavily in contact with the lunar profile, is the most uncertain, and (iii) according to the algorithm in Section 3.2, $A_1$ affects determinations of all $B_n$'s. The choice of the thickness of the layer has to be optimal; large enough to reduce uncertainties in $B_n$'s, but small enough to have a good resolution of the LDF. 
We chose $h=30$ mas for the lunar valley at AA = 177$^\circ$ and $h=73$ mas for the lunar valley at AA = 171$^\circ$.

\subsection{The Motion of the Solar Limb}

The motion of the Sun in the lunar valley is simulated with the Occult 4 software. 
If $\omega_{y}$ is the vertical velocity of the solar limb with respect to the lunar edge and $T$ is the duration of the light curve, every layer is filled in $t=h/\omega_{y}$ and the number of the layers is $N={\rm int}(T/t)$. 

As an example for the Nugent's video, from the bead at AA = 177$^\circ$ we obtain: $\omega_{y}=27.4$ mas s$^{-1}$, $t=1.11$ s, $T \approx 23.5$ s, and $N=21$.

\begin{figure}
\centerline{\includegraphics[width=0.9\textwidth,clip=]{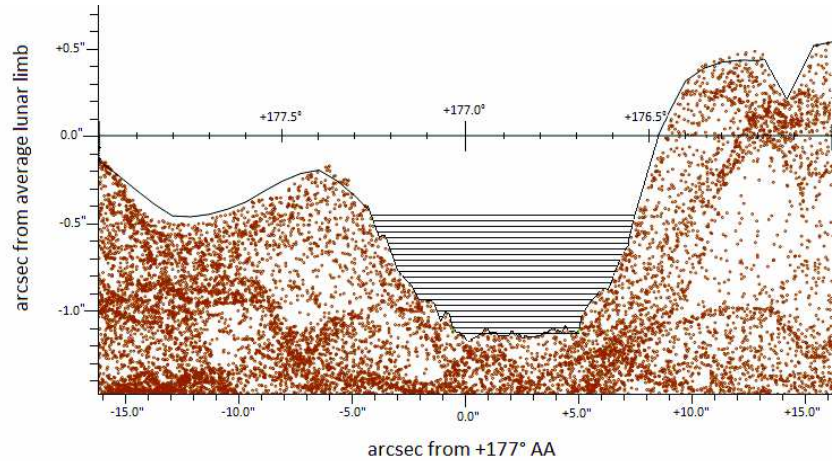}}
\caption{The lunar valley at AA=177$^{\circ}$ is divided into 21 layers. The thickness of each layer is 30 mas.}
\label{Fig. 9}
\end{figure}

\subsection{Results}

\begin{figure}
\centerline{\includegraphics[width=0.9\textwidth,clip=]{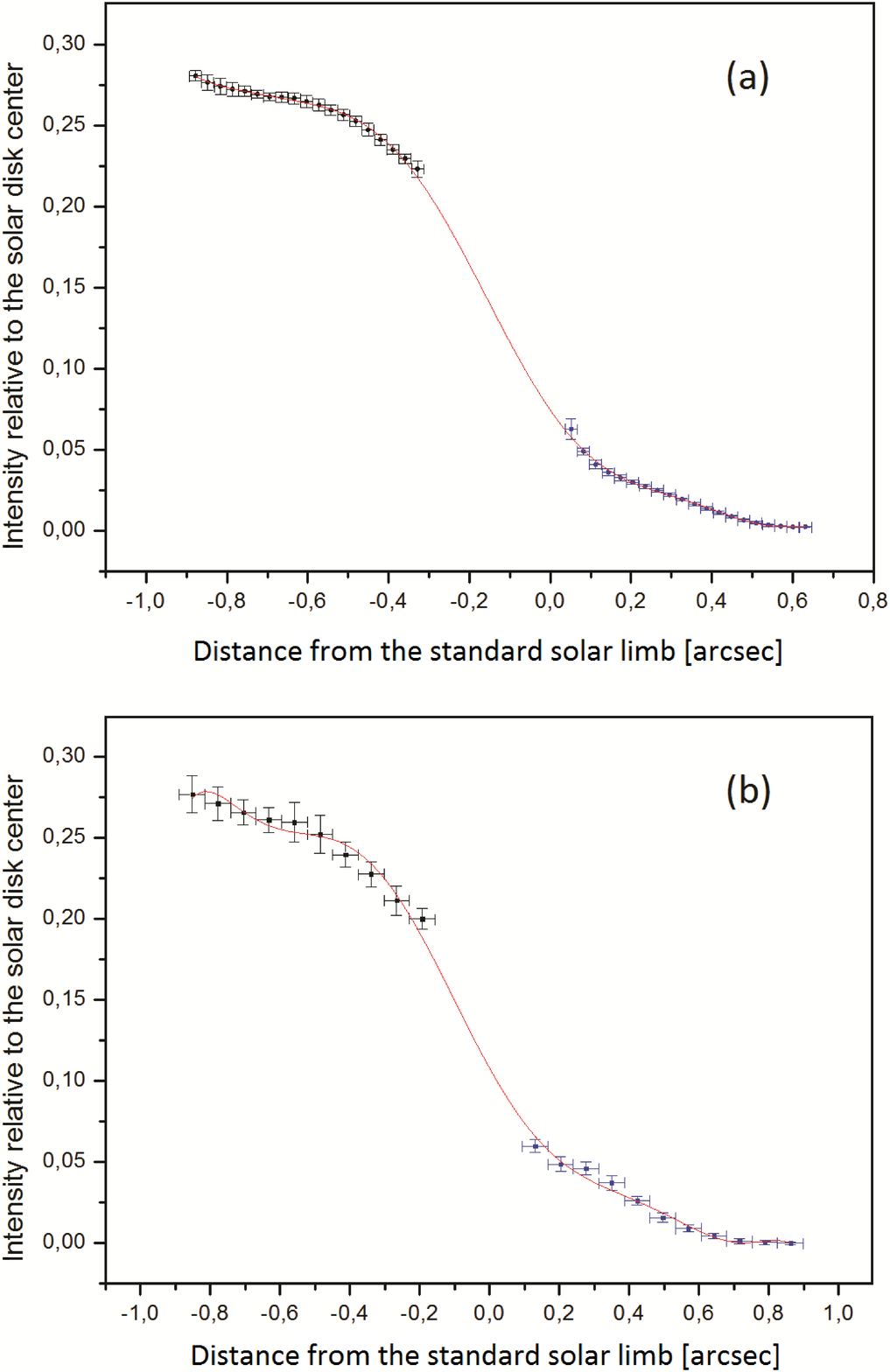}}
  
\caption{The luminosity profiles obtained are plotted and put together. The inner and brighter parts are obtained from Tegtmeier's video; the outer and weaker parts are obtained from Nugent's video. Panels (a) and (b) show respectively the profiles of the beads at AA = 177$^\circ$ and AA = 171$^\circ$. The luminosity profiles are normalized to the center of the solar disk according to \protect\inlinecite{Rogerson} for the inner parts, and in an arbitrary way for the outer parts. The zero of the abscissa is the position of the standard solar limb with a radius of 959.63 arcsec at 1 AU. The error bars on the $y$-axis are the 90\% confidence level. The error bars on the $x$-axis are the thickness ($h$) of the lunar layers. The solid line is an interpolation between the profiles and gives a possible scenario on the position of the inflection point.}
\label{Fig. 10}
\end{figure}

A program in Fortran90 is performed to calculate the points along the LDF ($B_m$'s).  The program takes into account the polynomial fits of the light curve and three values for each lunar layer area ($A$, $A+\Delta A$, $A-\Delta A$), giving a distribution of values for each point $B_n$. Figure 10 shows the results. 

The resulting points show that the inflection point is clearly between the two profiles obtained for each of the two beads. The saturation of the CCD pixels precluded measuring inward the luminosity function for Nugent's video, while low sensitivity made it impossible to measure outward the luminosity function for Tegtmeier's video. Therefore it is impossible to infer an exact location of the inflection point, but it is possible to deduce upper and lower limits to the location of the infection point: -0.19 arcsec $< \Delta R <$ +0.05 arcsec. 

The same eclipse was recorded by \inlinecite{Adassuriya}. The video analysis they performed with the classical approach (explained in Section 3.1) led to a value of $\Delta R = +0.26 \pm 0.18$ arcsec. This result does not seem compatible with the possible range for the position of the inflection point we found. This shows that the solar radius defined by the classical method is different from that defined by the inflection point of the LDF.

\section{Past and Future Observations}
The comparison of the shape and position of the LDF at different periods allows us to infer the behavior of the solar diameter over time. The method described above makes it possible to compare the results obtained with other methods of measurement of the LDF, as drift-scan or heliometric methods \cite{Sigi2011}; and of course the new method of the eclipse can be reproposed for future observations and also for the analysis of past eclipses.

\subsection{Analysis of Historical Eclipses} 
Historical observations of eclipses made with the naked eye could be important sources of information on the behavior of the solar diameter over long periods. Unfortunately these observations are affected by many uncertainties: human reaction time, uncertainty about the ephemeris, uncertainty about the time recording, and uncertainty on the coordinates of observation. However, there are some observations of eclipses which allow us to bypass most of these uncertainties. For example, the observation of an annular eclipse, where the totality was expected, allows us discussion on the phenomenon, regardless of timing, coordinates, and ephemeris.  This is the case for the observation of the annular eclipse made by the Jesuit astronomer Christopher Clavius on 9 April 1567 in Rome \cite{Clavius}. If the ring of that annular eclipse was a solar layer inner to the inflection point, the angular radius of the Sun would have been some arcsec larger than its standard value of 959.63 arcsec (see Figure 11), which is a surprising result. Taking this observation as a proof of a larger solar diameter, \inlinecite{Eddy} assumed a secular shrinking of the Sun from the 17th century to the present.

\begin{figure}
\centerline{\includegraphics[width=1.0\textwidth,clip=]{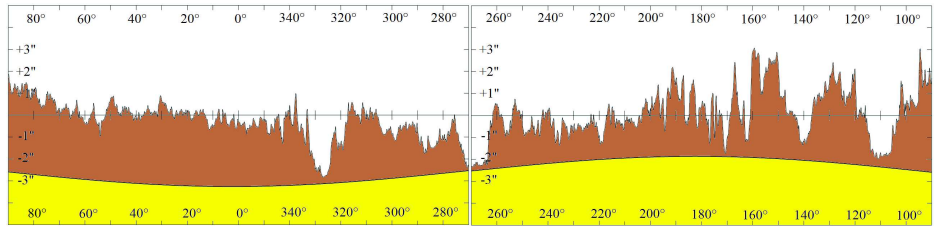}}
\caption{Eclipse of 9 April 1567 simulated with Occult the 4.0 software. A view from Collegio Romano (latitude = 41.90$^\circ$ N, longitude = 12.48$^\circ$ E) where probably he made his observation. The lunar edge's mountains are plotted as a function of the axis angle AA. 1$^\circ$ in the AA (abscissa) corresponds to 16 arcsec at the mean lunar distance. The solid line under the mountains is the standard solar limb. The curves are the northern (left) and the southern (right) semicircles.}
\label{Fig. 11}
\end{figure}

In the previous sections we discussed the incompatibility of the classical approach, compared to methods that take into account the inflection point of the LDF. The analysis of the eclipse observed by Clavius, with the classical approach, would lead to an angular diameter of the Sun greater than that of the Moon. But if we take into account the accepted definition of the solar limb we have to consider that the ring of the annular eclipse may have been external to the inflection point, and necessary corrections should be applied.

The detection of the LDF, in absolute magnitude, in the $V$ band, is thus an important future task for considering the effect of the outer regions of the solar atmosphere for naked eye observations. In particular there is a need to assess the distance from the inflection point where the light is no more visible for the naked eye, and the method proposed is well suited for this purpose. In this way we can get information about the position of the inflection point through the analysis of this and other historical eclipses.

\subsection{Recommendations for Future Observations}
Many possible improvements are possible to obtain the LDF with the method proposed:
\begin{itemize}
\item{An increased dynamic range of the CCD or CMOS detectors from 8 bits corresponding to 256 levels to 12 bits, which correspond to 4096 level of detectable intensity, is recommended. In this way we can extend the sampling of the luminosity function to regions of the photosphere more internal and luminous than the inflection point, without being saturated before it.} 
\item{A right setting of the sensitivity of the optical instruments is necessary to sample the luminosity function in regions outer and weaker than the inflection point.}
\item{As stated in Section 2.3, the profile is highly dependent on the wavelength. One must then observe at specific photometric bands to be able to compare the results with other measurements. The current space mission {\it Picard} is a good opportunity to compare the results. It is therefore recommended to observe the next eclipses at the same wavelengths at which the satellite {\it Picard} works (535, 607, 782 nm)\footnote[7]{\url{http://smsc.cnes.fr/PICARD/GP_instruments.htm}}.}
\item{The detection of the LDF in absolute magnitude is an important future task. The LDF profile is the best source of information for the solar atmosphere. The LDF profiles obtained from different wavelengths are useful to constrain the models \cite{Thuillier}.}
\end{itemize}

\section{Conclusions}

This study takes into account the potentiality of the observation of eclipses in defining the luminosity profile of the limb of the Sun. 
Huge developments in the method and means have been made, from the consideration of Baily's beads, to the recent mapping of the lunar surface by the satellite {\it Kaguya}. 

A further improvement takes place in this study, considering the bead as a light curve forged by the LDF and the profile of the lunar valley. A first application on two beads of the annular eclipse on 15 January 2010 is described in this study. We obtain a detailed profile, demonstrating the functionality of the method. Although it was impossible to observe the inflection point, its position is defined within a narrow range. The solar radius is thus defined within this range (from -0.19 to +0.05 arcsec with respect to the standard radius), which is consistent with the standard value.

Moreover, a consideration is coming out: the measurements with the classical approach depend on the optical instrument used. Hence, the classical method does not permit us to directly compare the values obtained with different instruments. This could be a good track for the investigation of the enigmatic eyewitnesses of the historical eclipses (the eye is the first ``optical instrument" used for observation of eclipses).

A variety of observation methods are employed to monitor the profile of the solar limb. In this contest the IOTA group could give an important contribution thanks to the new approach proposed. To exploit its potential in this paper we gave some recommendations for future observations.

\begin{acknowledgements}
This research has made use of the data taken with the LALT instrument on board the lunar orbiter {\it Kaguya} of JAXA/SELENE.
\end{acknowledgements}

\mbox{}~\\

\bibliography{LDF_SoPhys}
   
\bibliographystyle{spr-mp-sola}

\end{article}
\end{document}